\mag 1090
\tolerance 2000
\pretolerance 2000
\vsize=9.6truein \hsize=6.3truein
\lineskiplimit = 0pt
\parskip =2ex plus .5ex minus .1ex
\parindent 3em

\font\japbig = cmbx10 scaled \magstep1
\font\bbig = cmbx10 scaled \magstep2
\ifnum\mag=1090
 at 11truept
\font\japbig = cmbx10 at 12truept
\font\bbig = cmbx9 at 16truept
\font\bbbig = cmbx9 at 21truept
\fi
\newcount\equationnum
\global\equationnum=0
\def\bookdisp#1$${\line{\hfill{$\displaystyle#1$}
    \global\advance\equationnum by 1
    \hfill \llap{(\the\equationnum)}$\;$}$$}
\everydisplay{\bookdisp}

\centerline{\bbbig A diatribe on expanding space}

\medskip
\centerline{\bbig J.A. Peacock}
\centerline{\japbig Institute for Astronomy, University of Edinburgh}
\centerline{\japbig Royal Observatory, Edinburgh EH9 3HJ}

\bigskip

\noindent
This is an expansion of an analysis that 
first appeared in Peacock (2001), but which has not previously
been available online, except at {\tt www.roe.ac.uk/japwww}.
Some more details,
particularly analytic solutions for test-particle motion in open
and closed models, are given by Whiting (2004).
Some relevant further discussion is given by
Barnes et al. (2006).

\beginsection{\bbig 1 The meaning of an expanding universe}

The idea of an expanding universe
can easily lead to confusion, and this note tries to
counter some of the more tenacious misconceptions.
The worst of these is the `expanding space' fallacy. The
RW metric written in comoving coordinates emphasizes that
one can think of any given fundamental observer as fixed
in a coordinate system where separations increase 
in proportion to $R(t)$. A common
interpretation of this algebra is to say that the galaxies
separate ``because the space between them expands'', or
some such phrase. This seems a natural interpretation, but
we need to worry about what the coordinates mean,
as may be seen via two
examples: the empty universe and de Sitter space. In the
former case, Minkowski spacetime is rewritten as an expanding
open RW metric with $R(t)\propto t$. In the latter case, we
can compare the usual metric for de Sitter space
$$
c^2d\tau^2 = c^2dt^2 - R^2(t)\left[dr^2 + r^2 \,d\psi^2\right]; \quad R(t) \propto e^{Ht}
$$
with the static form in which de Sitter first derived it:
$$
c^2d\tau^2 = (1-r^2/{\cal R}^2)\,c^2dt^2 - 
(1-r^2/{\cal R}^2)^{-1} dr^2 - r^2 \,d\psi^2; \quad {\cal R}=c/H.
$$
It is not immediately obvious that there is anything expanding
about the second form, and historically this remained obscure
for some time. Although it was eventually concluded
(in 1923, by Weyl) that one would expect a redshift that
increased linearly with distance in de Sitter's model,
this was interpreted as measuring the constant radius of 
curvature of spacetime, ${\cal R}$. This is still the interpretation
given by Hubble in his 1929 attempt to detect the predicted
effect -- a paper that does not contain the word `expansion'.
But even if it takes more than just the appearance
of $R(t)$ in a metric to prove that something is expanding,
there are clearly cases where expansion is a legitimate global concept.
This is most clear-cut in the
case of closed universes, where
the total volume is
a well-defined quantity that increases with time, so undoubtedly
space is expanding in that case. 

But even if `expanding space' is a correct {\it global\/}
description of spacetime, boes the concept have a meaningful {\it local\/}
counterpart? Is the space in my bedroom expanding, and what would
this mean? Do we expect the Earth to recede from the Sun as the
space between them expands?
The very idea suggests
some completely new physical effect that is not covered by
Newtonian concepts. However, on scales much smaller than the current
horizon, we should be able to ignore curvature
and treat galaxy dynamics as occurring in Minkowski spacetime;
this approach works in deriving the Friedmann equation.
How do we relate this to `expanding space'? It should be
clear that Minkowski spacetime does not expand -- indeed, the
very idea that the motion of distant galaxies could affect
local dynamics is profoundly anti-relativistic: the equivalence
principle says that we can always find a tangent frame
in which physics is locally special relativity.

\beginsection{\bbig 2 Test-particle dynamics}

To clarify the issues here, it should help to consider an
explicit example, which makes quite a neat paradox.
Suppose we take a nearby low-redshift galaxy and give
it a velocity boost such that its redshift becomes
zero. At a later time, will the expansion of the
universe have caused the galaxy to recede from us, so that it
once again acquires a positive redshift? To idealize the
problem, imagine that the galaxy is a massless test particle
in a homogeneous universe.

The `expanding space' idea would suggest that the test
particle should indeed start to recede from us, and it appears that
one can prove this formally, as follows. Consider the
peculiar velocity with respect to the Hubble flow,
$\delta {\bf v}$. A completely general result is that this
declines in magnitude as the universe expands:
$$
\delta v \propto {1\over a(t)}.
$$
This is the same law that applies to photon energies, and
the common link is that it is particle momentum in general
that declines as $1/a$, just through the accumulated Lorentz
transforms required to overtake successively more distant
particles that are moving with the Hubble flow.
So, at $t\rightarrow\infty$, the peculiar velocity
tends to zero, leaving the particle moving with the Hubble
flow, however it started out: `expanding space' has apparently done
its job.

Now look at the same situation in a completely different way.
If the particle is nearby compared with the cosmological
horizon, a Newtonian analysis should be valid: in an
isotropic universe, Bikhoff's theorem assures us that we
can neglect the effect of all matter at distances greater
than that of the test particle, and all that counts is the
mass between the particle and us. Call the proper separation
of the particle from the origin $r$. Our initial conditions
are that $\dot r=0$ at $t=t_0$, when $r=r_0$. The equation
of motion is just
$$
\ddot r = {-G M(<r \mid t) \over r^2},
$$
and the mass internal to $r$ is just
$$
M(<r \mid t) = {4\pi\over 3}\, \rho r^3 = {4\pi\over 3}\, \rho_0\, a^{-3} r^3,
$$
where we assume $a_0=1$ and a matter-dominated universe.
The equation of motion can now be re-expressed as
$$
\ddot r = -{\Omega_0 H_0^2 \over 2 a^3}\, r.
$$
Adding vacuum energy is easy enough:
$$
\ddot r = -{H_0^2 \over 2 }\, r\, (\Omega_m a^{-3} -2 \Omega_v).
$$
The $-2$ in front of the vacuum contribution comes from the
effective mass density $\rho + 3p/c^2$.

We now show that this Newtonian equation is identical to what
is obtained from $\delta v \propto 1/a$. In our present notation,
this becomes
$$
\delta v = \dot r - H(t) r = -H_0 r_0 / a;
$$
the initial peculiar velocity is just $-Hr$, cancelling
the Hubble flow.
We can differentiate this equation to obtain $\ddot r$,
which involves $\dot H$. This can be obtained from the standard
relation
$$
H^2(t) = H_0^2 [\Omega_v + \Omega_m a^{-3} + (1-\Omega_m-\Omega_v)a^{-2} ].
$$
It is then a straightforward exercise to show that the equation
for $\ddot r$ is the same as obtained previously (remembering
$H=\dot a/a$).

Now for the paradox. It will suffice at first to solve the equation
for the case of the Einstein-de Sitter model, choosing
time units such that $t_0=1$, with $H_0 t_0=2/3$:
$$
\ddot r = -2 r /9 t^2.
$$
The acceleration is negative, so the particle moves {\it inwards\/},
in complete apparent contradiction to our `expanding space' conclusion that the
particle would tend with time to pick up the Hubble expansion.
The resolution of this contradiction comes from the full solution
of the equation. The differential equation clearly has power-law
solutions $r\propto t^{1/3}$ or $t^{2/3}$, and the combination with
the correct boundary conditions is
$$
r(t) = r_0 (2 t^{1/3} - t^{2/3}).
$$
At large $t$, this becomes $r = -r_0 t^{2/3}$. The use of a negative radius
may seem suspect, but we can regard $r$ as a Cartesian coordinate along
a line that passes through the origin, and the equation of motion
$\ddot r \propto r$ is correct for either sign of $r$. The
solution for $r(t)$ at large $t$ thus describes
a particle moving with the Hubble flow, but it arises because the particle has fallen
right through the origin and emerged on the other side.

In no sense, therefore, can `expanding space' be said to have
operated: in an Einstein-de Sitter model, a particle initially at rest
with respect to the origin falls towards the origin, passes through it,
and asymptotically regains its initial comoving radius on the
opposite side of the sky. The behaviour can be understood quantitatively
using only Newtonian dynamics.

This analysis demonstrates that there is no local effect on particle
dynamics from the global expansion of the universe: the tendency to separate
is a kinematic initial condition, and once this is removed, all memory of the expansion is lost.
Perhaps the cleanest illustration of the point is provided by the Swiss Cheese
universe, an exact model in which the mass within (non-overlapping) spherical
cavities is compressed to a black hole. Within the cavity, the metric is
exactly Schwarzschild, and the behaviour of the rest of the universe is
irrelevant. This avoids the small complication that arises when considering
test particles in a homogeneous universe, where we still have to consider the
gravitational effects of the matter between the particles.
It should now be clear how to deal with the question, ``does the expansion of
the universe cause the Earth and Moon to separate?'',
and that the answer is not the commonly-encountered
``it would do, if they weren't held together by gravity''.

Two further cases are worth considering. In an empty universe, the
equation of motion is $\ddot r=0$, so the particle remains at $r=r_0$,
while the universe expands linearly with $a\propto t$. In this
case, $H=1/t$, so that $\delta v = -Hr_0$, which declines as
$1/a$, as required.
Finally, models with vacuum energy are of more interest.
Provided $\Omega_v > \Omega_m/2$, $\ddot r$ is initially
positive, and the particle does move away from the origin.
This is the criterion for $q_0<0$ and an accelerating expansion.
In this case, there is a tendency for the particle to expand
away from the origin, and this is caused by the repulsive
effects of vacuum energy. In the limiting case of pure
de Sitter space ($\Omega_m=0$, $\Omega_v=1$), the particle's
trajectory is
$$
r = r_0 \cosh H_0 (t-t_0),
$$
which asymptotically approaches half the $r = r_0 \exp H_0 (t-t_0)$ that would
have applied if we had never perturbed the particle in the first place.
In the case of vacuum-dominated models, then,
the repulsive effects of vacuum energy cause all pairs of particles
to separate at large times, whatever their initial kinematics;
this behaviour could perhaps legitimately be called `expanding space'.
Nevertheless, the effect stems from the clear physical cause
of vacuum repulsion, and there is no new physical influence that arises
purely from the fact that the universe expands.
The earlier examples have proved that `expanding space' is in general
a dangerously flawed way of thinking about an expanding universe.

\beginsection{\bbig 3 The nature of the redshift}

Finally, some remarks about the relevance of the idea of
expanding space to the nature of the redshift.
For small redshifts, it is normal to interpret
the redshift as a Doppler shift ($z=v/c$). Even though
the idea of `expanding space' might challenge
such a view, it connects perfectly with the general
idea that $1+z$ measures the factor by which the
universe expanded between emission and absorption
of a photon.
Suppose we send a photon, which travels for
a time $\delta t$ until it meets another observer,
at distance $d=c\, \delta t$. The recessional
velocity of this galaxy is $\delta v = Hd$, so there
is a fractional redshift:
$$
\delta \nu \,/\, \nu = \delta v/c = - (Hd)/c = -H \delta t.
$$
Now, since $H=\dot R/R$, this becomes
$$
\delta \nu \,/\, \nu = -\delta R \, /\, R,
$$
which integrates to give the
main result: $\nu \propto 1/R$. As shown above,
the same reasoning proves that this $1/R$ scaling applies to the momentum
of all particles -- relativistic or not. Thinking
of quantum mechanics, the de Broglie
wavelength is $\lambda = 2\pi \hbar/p$, so this scales
with the side of the universe, yielding the common analogy of
standing waves trapped in an expanding box.

The redshift is thus the accumulation of
a series of infinitesimal Doppler shifts as the photon
passes from observer to observer, and this interpretation
holds rigorously even for $z\gg 1$.
However, this is not the
same as saying that the redshift tells us how fast the
observed galaxy is receding.
A common but incorrect approach is to use the
special-relativistic Doppler formula
and write
$$
1+z=\sqrt{{1+v/c\over 1-v/c}}.
$$
Indeed, it is all too common to read
of the latest high-redshift quasar as ``receding at 95\%
of the speed of light''. The reason the
redshift cannot be interpreted in this way is because
a non-zero mass density must cause gravitational
frequency shifts. Combining Doppler and gravitational
shifts, we then write
$$
1+z=\sqrt{1+v/c\over 1-v/c} \; \left(1+{\Delta\phi\over c^2}\right),
$$
where $\Delta\phi$ is the difference in gravitational potential
between the point of emission and reception of a photon.
If we think of the observer as lying at the centre of a sphere
of radius $r$, with the emitting galaxy on the edge, then
the sense of the gravitational shift will be a blueshift:
the radial acceleration at radius $r$ is $a=GM(<r)/r^2= 4\pi G\rho r/3$,
so the potential is thus $\Delta\phi=-4\pi G\rho r^2/6=-\Omega_mH_0^2 r^2/4$,
considering nonrelativistic matter only for simplicity.
The gravitational term is thus quadratic in $r$ and has to
be considered when going beyond first-order terms in the
Doppler shift. To second order, it
is exactly correct to think of the cosmological
redshift as a combination of doppler and gravitational
redshifts (see Bondi 1947 and problem 3.4 of 
`Cosmological Physics').

\def\japref{\parskip=0pt\par\noindent\hangindent\parindent
    \parskip =2ex plus .5ex minus .1ex}

\beginsection{\bbig References}

\japref
Barnes L.A., Francis M.J., James, J.B., Lewis G.F., 2006, 
MNRAS, 373, 382 (astro-ph/0609271)

\japref
Bondi H., 1947,  MNRAS, 107, 411

\japref 
Peacock J.A., 2001, in proc. 2000 Como School, eds
S. Bonometto, V. Gorini, U. Moschella [IOP], p9

\japref
Whiting A.B., 2004, Observatory, 124, 174 (astro-ph/0404095)

\bye